\documentstyle[preprint,prb,epsf,aps]{revtex}
\tighten
\begin{document}
\draft
\title{Mesoscopic Cooperative Emission from a Disordered System}
\author{T. V. Shahbazyan,$^1$ M. E. Raikh,$^2$ and Z. V. Vardeny$^2$}
\address{$^1$Department of Physics and Astronomy, 
Vanderbilt University, Box 1807-B, Nashville, TN 37235}
\address{$^2$Department of Physics, University of Utah, Salt Lake City,
UT 84112}
\maketitle

\begin{abstract}
We study theoretically the cooperative light  emission
from a system of $N\gg 1$ classical oscillators confined within a
volume  with spatial scale, $L$, much smaller than the radiation
wavelength, $\lambda_0=2\pi c/\omega_0$. We assume that the
oscillators frequencies are randomly distributed around a central
frequency, $\omega_0$, with some characteristic width,  
$\Omega\ll \omega_0$. In the absence of disorder, that is $\Omega=0$, 
the cooperative emission spectrum is composed of a narrow
subradiant peak superimposed on  a wide superradiant band. 
When $\Omega\neq 0$, we demonstrate that if $N$ is large
enough, the subradiant peak is not simply broadened by the disorder
but rather {\em splits into a  system of random narrow peaks}. We
estimate the spectral width of these peaks as a function of $N$, $L$,
$\Omega$, and $\lambda_0$. We also estimate the amplitude of this
mesoscopic structure in the emission spectrum.
\end{abstract}

\clearpage

\section{Introduction}
\label{sect:introduction}                                              
                                                                            
The study of cooperative phenomena in optics was initiated
by the pioneering work of Dicke.\cite{dicke}
The underlying physics of the cooperative emission can
be readily understood using  a classical approach.
Suppose, that a large number $N$ of identical oscillators
with frequency $\omega_0$ are confined within a small 
volume with characteristic size $L\ll 2\pi c/\omega_0=\lambda_0$,
where $\lambda_0$ is the radiation wavelengh; this is reffered to as
a ``point''  sample.\cite{rev} If $\tau$ is the radiative lifetime
of an isolated oscillator, then according to Dicke, the 
 $N$ eigenmodes   of the system of oscillators consist of one 
mode with a short lifetime $\tau/N\ll \tau$, and $N-1$ 
modes with lifetimes much longer than $\tau$ 
[by a factor $\sim (L/\lambda_0)^2$].
Correspondingly, the emission spectrum of this system
consists of superimposed broad (superradiant) and a narrow (subradiant)
bands. The intensities ratio of these bands is
  determined by the details of the excitation.
This type of lifetimes  redistribution   is caused by the interactions 
among the oscillators through their radiation fields.

Certainly, the classical picture does not describe 
all  aspects  of  the  cooperative emission. In fact, 
the original work of Dicke\cite{dicke} primarily
addressed  the time evolution of the radiation 
emission, provided that 
at the  initial moment, $t=0$,
 all the  oscillators are coherently excited.
For this situation, the classical picture helps in
understanding that  the radiation is 
released during a short time, $\tau/N$;  understanding
of the initial stages of the emission process
 (the delay time statistics)
requires, however, a quantum description.\cite{gross,andreev}
The original treatment in Ref. {\onlinecite{dicke}}
also ignored the dipole--dipole interactions, which 
give rise to a spread in the oscillators  frequencies (dephasing).
The question whether or not this dephasing would completely
destroy the cooperative emission is very non--trivial and was
addressed in a number of  later works.\cite{d-d,d-d',d-d''}  
 
In the previous  considerations of cooperative emission, it was
assumed that all $N$ oscillators (atoms, molecules or excitons) have
the same frequencies. Such a restriction was adequate for
the experimental
situation in both gases and single crystals. 
  To the best of our knowledge,
the only account  of  disorder
in the frequencies of the oscillators was given 
in Ref. \onlinecite{mukamel}, which addressed the transient behavior
 of the cooperative emission from  molecular aggregates.
The case of $J$--aggregates corresponds to a symmetrical
arrangement of oscillators in a circle. The authors\cite{mukamel}
treated the disorder within the perturbation theory and
averaged the second--order correction to the decay rates (the first--order
correction vanishes upon averaging) with a Gaussian distribution.
The  advantage of the work in Ref.\onlinecite{mukamel} is that the
nearest neighbors  dipole--dipole interactions   
were taken into account exactly. The drawback is in the perturbative
approach, which rules out certain ${\em qualitative}$ physical
effects (see below).

Whereas Ref.\onlinecite{mukamel} addressed a rather
particular situation, the following  {\em general} questions might
be asked. Suppose that the  oscillators frequencies  are
randomly distributed with a characteristic width $\Omega$.
Obviously, as $\Omega$ increases, it would eventually destroy
the cooperative features in the emission spectrum.
Then what is the critical magnitude of $\Omega$? How does
this magnitude depend on the parameters of
the system $N$, $L$, and $\lambda_0$? What is the structure
of the emission spectrum when disorder is smaller than critical? 

These questions have become not purely academic due to the recent 
advances in the field of laser--action in 
$\pi$--conjugated  polymers.\cite{tessl,hide,vard,varde,varden,vardeny,gel}
Some experiments provide a strong evidence for cooperative  emission 
from an ensemble of excitons in these materials, for
excitation intensities exceeding a certain characteristic
threshold.\cite{vard,varde,varden,vardeny,gel}
On the other hand, it is well known that the films of 
$\pi$-conjugated polymers are
strongly disordered\cite{hell} (in the absense of disorder,
cooperative emission by polymer chain was considered in 
Ref. \onlinecite{philpott}). They contain impurities and defects
which  break the polymer chains into segments of relatively short
conjugation length, with a  distribution depending on the film
quality.\cite{pich} This has a direct effect on the exciton energy,
$\hbar \omega$, since it has been found that $\hbar \omega$ directly
depends on the chain conjugation length.\cite{con}

The  questions formulated above are addressed in the present paper.
We study here the effect of disorder on the cooperative 
emission spectrum
of the system of classical oscillators. We consider the
situation of incoherent excitation, which is 
most relevant to the experiment.\cite{vard,varde,varden,vardeny,gel}
In contrast to Ref. \onlinecite{mukamel},
we are interested in the nonaveraged (but universal) properties of
the emission spectrum. In other words, our goal is to assess the
${\em mesoscopic}$\cite{mesos} aspects of the cooperative emission.
By mesoscopic we mean that, in the presence of a disorder, the
emission spectrum  of a large number of oscillators develops a fine
structure. The actual shape of this spectral
structure represents the 
{\em fingerprints} of the distribution of the  oscillator
frequencies and positions for a given realization. 
At the same time, the characteristic period and amplitude of the
fine structure are determined by the net parameters of the system:
$N$, $L$, and $\Omega$.

The paper is organized as follows. In Section\ \ref{basic} we derive the
expression for the emission spectrum of a system of classical
oscillators coupled by their radiation fields. In Section\ \ref{simple} we study in
detail a simplified model in which the coupling among the oscillators
is independent of distance.  The eigenmodes of 
a ``point'' sample in the presence of  disorder are analyzed in
Section\ \ref{structure}. The universal properties of the mesoscopic
stucture in the emission spectrum for small and large (but still
smaller than $\lambda_0$) sizes $L$ are discussed in 
Sections\ \ref{mesoscopic} and \ref{scaling}, respectively.
In Section\  \ref{discussion} the effect of the dipole--dipole
interactions is addressed. The conclusions are given in 
Section\  \ref{conclusion}. 

\section{The Basic Equations}
\label{basic}                                              

We consider a system of $N$ oscillators 
located at random points ${\bf r}_{i}$, with frequencies $\omega_i$
randomly distributed around a central frequency  $\omega_0$ with a
characteristic width $\Omega$. Each oscillator is driven  by the
radiation field ${\bf E}({\bf r},t)$ produced by all oscillators.
The equation of motion for the displacement $u_i$ of a given
oscillator $i$  reads 
\begin{equation}
\label{dip}
\ddot{u}_i + \omega_i^2  u_i = 
\frac{e}{m}\, {\bf n}_i\cdot {\bf E}({\bf r}_i,t),
\end{equation}
where $e$ and $m$ are the dipole characteristics (effective charge
and mass) and ${\bf n}_i$ is a unit vector in the direction of the
dipole moment. 

The current density, associated with the oscillators  motion, can
be written as
\begin{equation}
\label{density}
{\bf J}({\bf r},t)= e\sum_i {\bf n}_i \dot{u}_i\delta({\bf r}-{\bf r}_i).
\end{equation}
The current ${\bf J}$ plays the role of a source, which generates
the electric field ${\bf E}({\bf r},t)$ according to
\begin{equation}
\label{ex}
\Delta{\bf E}-\frac{1}{c^2}\ddot{\bf E}=\frac{4\pi}{c^2}\dot{\bf J},
\end{equation}
where $c$ is the speed of light.

Within the classical approach, the emission spectrum of the system 
should be calculated as follows. We assume that at the initial moment,
$t=0$, all oscillators are excited with  different phases $\phi_i$,
and that the radiation field 
at the  initial moment is zero, ${\bf E}({\bf r}, 0)=0$.
The evolution of ${\bf E}$ with time can be then obtained by  solving
Eqs.\ (\ref{dip})--(\ref{ex}). After taking the limit 
$r\rightarrow \infty$ and expanding the field into harmonics, the 
spectral intensity can be obtained as 
$I(\omega)=|{\bf E}(\infty, \omega)|^2$. 

To carry out this program, it is convenient to employ the Laplace
transformation. The transformed functions $\bar{u}_i(p)$
and $\bar{\bf E}({\bf r},p)$ satisfy the following system of equations
\begin{eqnarray}
(\omega_i^2 +p^2)\bar{u}_i(p)&=&\frac{e}{m}\,
{\bf n}_i\cdot\bar{\bf E}({\bf r}_i,p)
+u_0(p\cos\phi_i -\omega_i\sin\phi_i),\label{trans}
\\
\Delta \bar{\bf E}({\bf r},p)
-\frac{p^2}{c^2}\,\bar{\bf E}({\bf r},p)
&=&\frac{4\pi e}{c^2}\sum_i {\bf n}_i 
\left[p^2 \bar{u}_i(p)-u_0(p\cos\phi_i 
-\omega_i\sin\phi_i)\right]\delta({\bf r}-{\bf r}_i),\label{J}
\end{eqnarray}
where $u_0\sin\phi_i$ and $\omega_i u_0\cos\phi_i$ are the respective 
initial
displacement and  velocity  of the  $i$th oscillator.
The solution of Eq.\ (\ref{J}) for $\bar{\bf E}({\bf r},p)$ can be
presented as a superposition of eigenmodes,
  ${\bf E}_{\nu}({\bf r})$, of
the wave equation for the electromagnetic field,
\begin{equation}
\label{E}
\Delta {\bf E}_{\nu}({\bf r})+
\frac{\omega_{\nu}^2}{c^2}{\bf E}_{\nu}({\bf r})=0,
\end{equation}
where $\omega_{\nu}$ is the eigenfrequency.
Assuming that the modes are normalized, 
$\int d{\bf r} {\bf E}_{\nu}^2({\bf r})=1$,
we  obtain the following expression for $\bar{\bf E}({\bf r},p)$
\begin{equation}
\label{expa}
\bar{\bf E}({\bf r},p)=-4\pi e\sum_{i\nu}
\left[p^2 \bar{u}_i(p)-u_0(p\cos\phi_i -\omega_i\sin\phi_i)\right]
\frac{{\bf n}_i\cdot {\bf E}_{\nu}({\bf r}_i)}
{\omega_{\nu}^2 + p^2}\,{\bf E}_{\nu}({\bf r}).
\end{equation}
Substituting Eq.\ (\ref{expa}) into Eq.\ (\ref{trans}), we get a system
of coupled equations for the amplitudes $\bar{u}_i(p)$ 
\begin{eqnarray}
\label{sys}
(\omega_i^2 +p^2)\bar{u}_i(p)=
-\frac{4\pi e^2}{m}\sum_{j\nu}
\frac{
\left[{\bf n}_i\cdot {\bf E}_{\nu}({\bf r}_i)\right]
\left[{\bf n}_j\cdot {\bf E}_{\nu}({\bf r}_j)\right]
}
{\omega_{\nu}^2+ p^2}
&&
\left[p^2 \bar{u}_j(p)-u_0(p\cos\phi_j -\omega_j\sin\phi_j)\right]
\nonumber\\ &&
+u_0(p\cos\phi_i -\omega_i\sin\phi_i).
\end{eqnarray}
To simplify Eq.\ (\ref{sys}), it is convenient to introduce new
variables $v_i(p)$: 
\begin{equation}
\label{v}
v_i(p)=\frac{p^2\bar{u}_i(p)}{u_0}-(p\cos\phi_i -\omega_i\sin\phi_i).
\end{equation}
Then Eq.\ (\ref{sys}) takes the form
\begin{equation}
\label{newsys}
(\omega_i^2 + p^2)v_i+\sum_j S_{ij}v_j = 
\omega_i^2(\omega_i\sin\phi_i - p\cos\phi_i),
\end{equation}
where the coefficients 
\begin{equation}
\label{int}
S_{ij}(p)=\frac{4\pi e^2p^2}{m}\sum_{\nu}
\frac{
\left[{\bf n}_i\cdot {\bf E}_{\nu}({\bf r}_i)\right]
\left[{\bf n}_j\cdot {\bf E}_{\nu}({\bf r}_j)\right]
}
{\omega_{\nu}^2+p^2}
\end{equation}
describe the coupling between oscillators $i$ and $j$ via the
radiation field. 

Let us now express the intensity, $I(\omega)$, in terms of the
variables $v_i(p)$.  
The expression for $\bar{\bf E}({\bf r},p)$ follows from 
Eqs.\ (\ref{expa}) and $(\ref{v})$
\begin{equation}
\label{fie}
\bar{\bf E}({\bf r},p)=-4\pi e\sum_{i\nu}v_i(p)
\frac{{\bf n}_i\cdot {\bf E}_{\nu}({\bf r}_i)}
{\omega_{\nu}^2 + p^2}{\bf E}_{\nu}({\bf r}).
\end{equation}
The Fourier transform of the electric field is obtained by replacing
 $p$ by $i\omega$ in Eq.\ (\ref{fie}). In the limit 
$r\rightarrow \infty$, only the pole $\omega_{\nu}=\omega$
 contributes to the sum over $\nu$, so that
\begin{equation}
\label{four}
{\bf E}({\bf r},\omega)|_{r\rightarrow\infty}\propto
\sum_{i\nu} v_i\left[{\bf n}_i\cdot {\bf E}_{\nu}({\bf r}_i)\right]
{\bf E}_{\nu}({\bf r})\delta (\omega_{\nu}^2-\omega^2).
\end{equation}
This corresponds to taking the continuum limit for electromagnetic modes.
The terms proportional to
$\left[{\bf E}_{\nu}({\bf r})\cdot {\bf E}_{\mu}({\bf r})\right]$,
which appear after calculating $|{\bf E}({\bf r},\omega)|^2$ from 
Eq.\ (\ref{four}), oscillate rapidly  if $\mu \neq \nu$. Therefore,
only the terms with $\mu=\nu$ survive at large $r$. These terms
contain products of the form 
$\left[{\bf n}_i\cdot {\bf E}_{\nu}({\bf r}_i)\right]
\left[{\bf n}_j\cdot {\bf E}_{\nu}({\bf r}_j)\right]$.
Note that the same products enter into the coupling 
coefficients, $S_{ij}$, defined by Eq.\ (\ref{int}). This allows us to present the
final expression for the spectral intensity in a compact form
\begin{equation}
\label{I}
I(\omega)\propto\sum_{ij}v_i({\rm Im}S_{ij})v_j^{\ast},
\end{equation}
where $v_i$ and $S_{ij}$ are calculated at $p=i\omega$.

We assume that the  spread of the  oscillators frequencies due to
the disorder is much smaller than the central frequency,
$\Omega\ll\omega_0$. This means that the frequency dependence of the
coupling constants is weak, so that $S_{ij}(i\omega)$ can be
evaluated at $\omega=\omega_0$.  The real part of $S_{ij}$, which
comes from the principle value of the sum  over modes in Eq.\
(\ref{int}), diverges for $i=j$. This divergency is the
manifestation of the Lamb shift, well--known in  quantum
electrodynamics, and can be absorbed into $\omega_i$. At the same
time, the imaginary  part of $S_{ii}$, which  results from the pole
$\omega_{\nu}=\omega_0$, is finite. It determines the radiative
lifetime, $\tau$, of an individual oscillator via the relation 
${\rm Im} S_{ii}(i\omega_0)=2\omega_0/\tau$. For a single
oscillator in  vacuum, the modes ${\bf E}_{\nu}$ are simply plane
waves, and the summation over $\nu$ in Eq.\ (\ref{int}) recovers the
textbook result 
\begin{equation}
\label{tau}
\tau=\frac{3mc^3}{e^2\omega_0^2}.
\end{equation}
For $i\neq j$, the coupling $S_{ij}$ between two oscillators depends 
on the ratio $r_{ij}/\lambda_0$, where $r_{ij}$ is the distance
between the oscillators, and $\lambda_0 = 2\pi c/\omega_0$ is the 
radiation wavelength. For $r_{ij}\gg\lambda_0$, both real and
imaginary parts of $S_{ij}$ oscillate rapidly with $r_{ij}$, and the
effect of coupling is negligibly small for a large ensemble of
oscillators.\cite{keller} For $r_{ij}\ll\lambda_0$, the real part of
$S_{ij}$  represents the dipole--dipole interaction of the 
oscillators $i$ and $j$. It is convenient to present $S_{ij}$
in the form 
\begin{equation}
\label{def}
S_{ij}=\frac{2\omega_0}{\tau}(\beta_{ij}+i\alpha_{ij}),
\end{equation}
where $\beta_{ij}$ and $\alpha_{ij}$ are the dimensionless matrices of
coupling between the oscillators, defined as 
\begin{equation}
\label{beta}
\beta_{ij}=\left(\frac{\lambda_0}{2\pi r_{ij}}\right)^3
\left[({\bf n}_i\cdot {\bf n}_j)-
\frac{3({\bf n}_i\cdot {\bf r}_{ij})
({\bf n}_j\cdot {\bf r}_{ij})}{r_{ij}^2}\right],
\end{equation}
and
\begin{eqnarray}
\label{alpha}
\alpha_{ij}={\bf n}_i\cdot {\bf n}_j
-\frac{1}{5}\left(\frac{2\pi r_{ij}}{\lambda_0}\right)^2
\left[({\bf n}_i\cdot {\bf n}_j)-
\frac{({\bf n}_i\cdot {\bf r}_{ij})
({\bf n}_j\cdot {\bf r}_{ij})}{2r_{ij}^2}\right].
\end{eqnarray}
Turning back to Eq.\ (\ref{newsys}), we note that since the
distribution of  oscillators frequencies  is  
relatively narrow, that is $\Omega \ll \omega_0$, we can make some
simplifications. Namely, for $p=i\omega$, the
factor $(\omega_i^2+p^2)$ in the lhs can be replaced by
$2\omega_0(\omega_i-\omega)$, 
and the rhs can be written as $-i\omega_0^3 e^{-i\phi_i}$.
Finally, after rescaling $v_i$ by factor $\omega_0^2$, 
Eq.\ (\ref{newsys}) takes the form
\begin{equation}
\label{lastsys}
(\omega_i-\omega)v_i+\frac{1}{\tau}\sum_j(\beta_{ij}+i\alpha_{ij})v_j
=-\frac{i}{2}e^{-i\phi_i}.
\end{equation}
Equation (\ref{lastsys}) together with Eqs.\ (\ref{I}) and 
(\ref{def})--(\ref{alpha}) allow us to calculate
the spectral intensity $I(\omega)$ for any set of initial
oscillators phases. For $\alpha_{ij}=\beta_{ij}=0$ ($i\neq j$), the
eigenfrequencies of the system are simply the frequencies of individual
oscillators, and the the emission spectum represents a superposition of
Lorentzian peaks centered at $\omega_i$. In the presence of
nondiagonal coupling, the eigenfrequencies are those of
{\em cooperative eigenmodes} which, in turn, are determined by the
imaginary part of the coupling, $\alpha_{ij}$.
In the experiment, the measured spectrum
represents the result of averaging over many 
excitation pulses.\cite{varden} In
order to simulate the experimental situation, we will assume the
phases $\phi_i$ to be  uncorrelated random numbers  and  average the
result for the spectral intensity over all $\phi_i$.

\section{A Simple Model}
\label{simple}

In this section  we consider a simplified situation, in which 
Eq.\ (\ref{lastsys}) with random  frequencies $\omega_i$ can be
solved exactly and the expression for the spectral intensity can be
obtained in a closed form. Following Dicke,\cite{dicke} we
disregard the dipole--dipole interactions by setting $\beta_{ij}=0$. 
Although this approximation is rather common, later on we will 
discuss it in more detail. Turning to $\alpha_{ij}$, we note that since 
$L^2/\lambda_0^2\ll 1$, the second term in Eq.\ (\ref{alpha}) is a
small correction to the first term. We therefore approximate  the
{\em non--diagonal} elements of $\alpha_{ij}$ by
replacing $r_{ij}^2/\lambda_0^2$ with its average,
\begin{equation}
\label{separab}
\alpha_{ij}=\alpha {\bf n}_i\cdot {\bf n}_j,
\,\,\,\,\,\,\alpha_{ii} =1,
\end{equation}
where the coupling constant $\alpha$, with a typical value 
$(1-\alpha)\sim L^2/\lambda_0^2\ll 1$, is the same for {\em all} pairs. 
Note however, that the disorder coming
from random orientations of  ${\bf n}_i$ is still included. 
Later we will use this model for the analysis
of the system (\ref{lastsys}) with realistic $\alpha_{ij}$.

\subsection{General solution}

For the model coupling (20), the system of equations (\ref{lastsys}) takes the form
\begin{equation}
\label{manyvec}
\left[\omega_i-\omega + \frac{i}{\tau}(1-\alpha)\right]v_i
+\frac{i}{\tau}\alpha {\bf n}_i\cdot{\bf s}=-\frac{i}{2}e^{-i\phi_{i}},
\end{equation}
with vector ${\bf s}$ defined as
\begin{equation}
\label{svec}
{\bf s}=\sum_{i=1}^N v_i {\bf n}_i.
\end{equation}
A closed equation for ${\bf s}$
can be obtained by  multiplying $v_i$, found from 
Eq.\ (\ref{manyvec}), by ${\bf n}_i$ and
taking the sum over $i$. This yields
\begin{equation}
\label{for s}
{\bf s}+\frac{i\alpha}{\tau}
\sum_i\frac{{\bf n}_i({\bf n}_i\cdot {\bf s})}
{\omega_i-\omega +i(1-\alpha)/\tau}
=-\frac{i}{2}\sum_i 
\frac{{\bf n}_i e^{-i\phi_i}} {\omega_i-\omega +i(1-\alpha)/\tau}.
\end{equation}
Solving  Eqs.\ (\ref{manyvec})\ and\  (\ref{for s}) for
$v_i$ and substituting the result into Eq.\ (\ref{I}), we obtain for
the spectral intensity after some algebra 
\begin{equation}
\label{Inonav}
I(\omega)\propto -{\rm Im}\left[f(\omega)
-\frac{i\alpha}{\tau}\sum_{\mu\nu}g_{\mu}^{-}
\left(1+\frac{i\alpha}{\tau} F\right)_{\mu\nu}^{-1}g_{\nu}^{+}\right],
\end{equation}
where we introduced a function 
\begin{equation}
\label{f}
f(\omega)=\sum_i\frac{1}{\omega_i-\omega+i(1-\alpha)/\tau},
\end{equation}
a vector 
\begin{equation}
\label{gmu}
g_{\mu}^{\pm}(\omega)=\sum_i 
\frac{e^{\pm i\phi_i}n_{i\mu}}{\omega_i-\omega +i(1-\alpha)/\tau},
\end{equation}
and a tensor
\begin{equation}
\label{Fmunu}
F_{\mu\nu}(\omega)=\sum_{i}
\frac{n_{i\mu}n_{i\nu}} {\omega_i-\omega +i(1-\alpha)/\tau},
\end{equation}
where $n_{i\mu}$ are the components of ${\bf n}_i$. 

\subsection{Identical oscillators}

Let us first consider the case of $N$ identical oscillators having
the same frequencies $\omega_i=\omega_0$, and dipole momenta
all aligned in 
the same direction. Then we find from Eq.\ (\ref{Fmunu}),
\begin{equation}
\label{ideal F}
F_{\mu\nu}(\omega)=\delta_{\mu\nu}f(\omega)
=\frac{\delta_{\mu\nu}N}{\omega_i-\omega +i(1-\alpha)/\tau},
\end{equation}
and after averaging over the initial phases $\phi_i$, we obtain from
Eq.\ (\ref{Inonav})
\begin{eqnarray}
\label{N}
I(\omega)\propto \left[\frac{(N-1)(1-\alpha)/\tau}
{(\omega_0-\omega)^2+(1-\alpha)^2/\tau^2}+
\frac{(1-\alpha +\alpha N)/\tau}
{(\omega_0-\omega)^2+(1-\alpha +\alpha N)^2/\tau^2}\right].
\end{eqnarray}
The emission spectrum is a superposition of a
wide and a narrow Lorentzians with spectral widths 
$\Gamma\simeq N/\tau$ and $\gamma=(1-\alpha)/\tau$, respectively.
In accordance to the classical
result,\cite{dicke} the eigenmodes of the system of $N$ identical
oscillators coupled via their radiation field represent a single
superradiant mode with short radiation time $\tau/N$,
and $N-1$ subradiant modes with radiation time much longer than that
for an isolated oscillator, $\tau/(1-\alpha)\gg \tau$.
The superradiant mode is a symmetric superposition of oscillator
states and is strongly coupled to the radiation field, whereas the
coupling of the subradiant modes to the radiation field is suppressed. 
In this case, the frequencies of all $N-1$ subradiant modes are
degenerate, and the spectrum consists of a single narrow peak of
width $\gamma$ on top of much broader band of width $\Gamma$, as
shown in Fig.\ 1. As can be seen, with decreasing $1-\alpha$, the
hight of the subradiant peak increases, whereas the amplitude of the
superradiand band diminishes.

\subsection{Random frequencies}

Consider now the case when the   oscillators frequencies are
random, but orientational disorder is still absent, i.e. all dipoles
are aligned in one direction. Again we have 
$F_{\mu\nu}(\omega)=\delta_{\mu\nu}f(\omega)$, with
$f(\omega)=f'(\omega)+if''(\omega)$ 
given by Eq.\ (\ref{f}). Then a straightforward evaluation of 
Eq.\ (\ref{Inonav}) yields (after averaging over the phases) 
\begin{eqnarray}
\label{Iav}
I(\omega)\propto
\left[
\frac{\frac{\alpha}{\tau}f'_1
\left(1-\frac{\alpha}{\tau}f''\right)
+\left(\frac{\alpha}{\tau}\right)^2f''_1 f'}
{\left(1-\frac{\alpha}{\tau}f''\right)^2+ 
\left(\frac{\alpha}{\tau}f'\right)^2}
-f''
\right].
\end{eqnarray}
where the function $f_1(\omega)=f'_1(\omega)+if''_1(\omega)$ is defined
as

\begin{equation}
\label{f_1}
f_1(\omega)=\sum_i\frac{1}{\left[\omega_i-\omega+i(1-\alpha)/\tau\right]^2}.
\end{equation}
In order to clarify the underlying
physics, it is useful to express the spectral intensity 
in terms of the system eigenmodes. The eigenfrequencies
$\tilde{\omega}_k$ are determined by the equation:
\begin{equation}
\label{eigenmod}
1+\frac{i\alpha}{\tau}f(\tilde{\omega}_k)=0.
\end{equation}
Then the 
intensity Eq.\ (\ref{Iav}) can be simply rewritten  as  
\begin{equation}
\label{Ieigen}
I(\omega)\propto \sum_k \frac{\tilde{\omega}''_k}
{(\omega-\tilde{\omega}'_k)^2+\tilde{\omega}''^2_k},
\end{equation}
where $\tilde{\omega}'_k={\rm Re}\,\tilde{\omega}_k$ is the 
 eigenmode frequency
 and $\tilde{\omega}''_k={\rm Im}\,\tilde{\omega}_k$
characterizes its width. Note that for $\omega_i=\omega_0$, we have $N-1$
degenerate eigenmodes with $\tilde{\omega}'_k=\omega_0$, and 
Eq.\ (\ref{Ieigen}) turns into Eq.\ (\ref{N}).  

\subsection{Disorder in orientations}

In the presence of the orientational disorder, the
spectral intensity (\ref{Inonav}) depends, in principle, 
on the direction of each ${\bf n}_i$. However, for large $N$, 
one can replace  the product $n_{i\mu}n_{i\nu}$ in 
Eq.\ (\ref{Fmunu}) for $F_{\mu\nu}$, with its average,
\begin{equation}
\label{orient}
\langle n_{i\mu}n_{i\nu}\rangle= \frac{1}{3}\delta_{\mu\nu}.
\end{equation}
Thus, we have
$F_{\mu\nu}(\omega)=\frac{1}{3}\delta_{\mu\nu}f(\omega)$, so that the 
expression for the spectral intensity is similar 
to Eq.\ (\ref{Iav})  with the only difference that in the first term,
the functions $f(\omega)$ and $f_1(\omega)$ are now multiplied by $1/3$.
This results in a shrinkage of the superradiant emission band
 by the same factor. At the same time, the width of subradiant peak
increases by a factor of 3. Thus, the orientational disorder has 
{\em no} qualitative effect on the cooperative emission
spectrum. The reason is that the coupling (20) is {\em separable}, 
that is it depends on orientations 
via the product ${\bf n}_i\cdot {\bf n}_j$. Furthermore, for realistic
$\alpha_{ij}$ given by Eq.\ (\ref{alpha}), the main (first) term has
the same separable form; therefore, the orientational disorder does
not qualitatively affect the  cooperative emission spectrum and will be
disregarded in the rest of the paper.

\subsection{Numerical results}

In Fig.\ 2 we plot the normalized spectral intensity in the absence of
coupling, i.e. $\alpha=0$, with  increasing number of oscillators. 
Each spectrum is calculated with a computer generated  set of $N$
random frequencies $\omega_i$, which we have chosen, for simplicity,
to be uniformly distributed in the interval
$(\omega_0-\Omega,\omega_0+\Omega)$. 
For convenience, the spectra corresponding to different $N$ are
normalized and shifted in the vertical direction. It can be seen
that the peaks are resolved in the spectrum  as long as the
disorder, $\Omega$, is larger than  $N/\tau$. We also see that for
sufficiently large $N$, the intensity peaks are  washed out from the
spectrum. 

In Figs.\ 3--6 we present the results for $I(\omega)$ calculated 
using  Eq.\ (\ref{Iav})  for several values of $\alpha$ close to 1. 
The striking feature of the emission spectrum is its 
{\em mesoscopic} character. In the presence of disorder, the narrow
subradiant peak of Eq.\ (\ref{N}) (see Fig. 1) is not smeared out
due to a large spread in $\omega_i$, as in the case of uncoupled
oscillators (see Fig.\ 2), but rather splits into 
{\em a multitude of narrow peaks} 
corresponding to the eigenmodes of the disordered system. 
Furthermore, although the curves are calculated with different
random sets of frequencies, the overall pattern of
the emission spectrum exhibits certain universal features. In
particular, it can be seen by comparing Figs.\ 3--6 that with
increasing $N$, the random structure of the spectrum  undergoes
several transformations, and that the characteristic $N$,  at which
the changes in the pattern occur, is sensitive to the proximity of
$\alpha$ to 1. This indicates a rather non--trivial structure of the
eigenmodes, which we address in the next section. 

\section{Structure of eigenmodes}
\label{structure}

The eigenmodes of a system of $N$ oscillators coupled through their
radiation field are determined by the homogeneous part of 
Eq.\ (\ref{lastsys})  (we set $\beta_{ij}=0$ in this section)
\begin{equation}
\label{eigen}
(\omega_i-\omega)v_i+\frac{i}{\tau}\sum_j \alpha_{ij}v_j=0.
\end{equation}
Since the typical values of  $(1-\alpha_{ij})\sim r_{ij}^2/\lambda_0^2$ 
are small, we split the second term in Eq.\ (\ref{eigen}) into a 
sum of the main contribution, with $\alpha_{ij}=1$, and a correction
proportional to $(\alpha_{ij}-1)$.
Analogously to the consideration in the previous section, we rewrite 
Eq.\ (\ref{eigen}) as  
\begin{equation}
\label{eigen1}
(\omega_i-\omega)v_i+\frac{i}{\tau} s (1+\sigma_i)=0,
\end{equation}
with 
\begin{equation}
\label{sigma}
s=\sum _j v_j,\,\,\,\,\,\,\, 
\sigma_i=\frac{1}{s}\sum_j (\alpha_{ij}-1)v_j.
\end{equation}
Expressing $v_i$ from Eq.\ (\ref{eigen1}) and taking the sum over
$i$, we obtain 
\begin{equation}
\label{eqeigen}
1+\frac{i}{\tau}\sum_j\frac{1+\sigma_j}{\omega_j-\omega}=0.
\end{equation}
The equation for $\sigma_i$ follows from substituting of $v_j$, 
found from Eq.\ (\ref{eigen1}), into the definition of $\sigma_i$,
Eq.\ (\ref{sigma}), 
\begin{equation}
\label{eqsigma}
\sigma_i+\frac{i}{\tau}\sum_j
\frac{\alpha_{ij}-1}{\omega_j-\omega}(1+\sigma_j)=0.
\end{equation}
The solutions of Eqs.\ (\ref{eqeigen})\ and\  (\ref{eqsigma})
determine the complex frequencies of the eigenmodes,
$\tilde{\omega}_k\equiv\tilde{\omega}'_k+i\tilde{\omega}''_k$.

\subsection{``Point'' sample}

Let us first analyze the effect of disorder on a  system
with all $\alpha_{ij}=1$, corresponding to the limit of a ``point''
sample, i.e.
$(L/\lambda_0)^2\ll 1$. With  $\sigma_i=0$, the real and
imaginary parts of  Eq.\ (\ref{eqeigen}) read
\begin{eqnarray}
\label{eigensys}
\frac{1}{\tau}\sum_j\frac{\omega_j-\omega'}
{(\omega_j-\omega')^2+\omega''^2}=0,\label{resys}
\\
\frac{1}{\tau}\sum_j\frac{\omega''}
{(\omega_j-\omega')^2+\omega''^2}=1.\label{imsys}
\end{eqnarray}
This system of equations has two different solutions with a
crossover between  them governed by the parameter $\Omega\tau/N$. 
For large disorder, $\Omega \gg N/\tau$, 
it can be readily seen that only one term in each of
Eqs. (\ref{resys}) and (\ref{imsys})  contributes
to the sum. In this case, the solutions are simply
$\omega=\omega_j+i/\tau$, as if the oscillators were uncoupled.
In fact, this conclusion could be anticipated. The above parameter
represents the ratio of the mean frequency spacing (MFS)
of oscillators, $\Omega/N$, and the inverse
lifetime of an individual oscillator, $1/\tau$; when the former is 
much larger than the latter, $\Omega/N \gg 1/\tau$, the
oscillators do not ``feel'' each other. 

In the opposite case of large $N$ (or weak disorder), $N\gg
\Omega\tau$, the analysis of Eqs. (\ref{resys}) and (\ref{imsys})
is carried out as follows. First note that in Eq.\ (\ref{resys}),
which determines the real parts of the eigenfrequencies,
$\tilde{\omega}'_k$, all the terms in the sum  contribute now. Let
us  drop $\omega''^2$ in the denominator of Eq.\ (\ref{resys}) (this
step will be justified below). Then we obtain that the solutions 
$\tilde{\omega}'_k$ are given  by the extrema of the polynomial
$P(\omega)=\prod_j(\omega_j-\omega)$.
These determine the frequencies of the $N-1$ {\em subradiant}
modes. At the same time, in Eq.\ (\ref{imsys}), which
determines the imaginary parts of the eigenfrequencies,
$\tilde{\omega}''_k$, all the terms in the sum are positive,
so that one should keep only the term with $\omega_j$ closest to
$\tilde{\omega}'_k$. Since
$(\tilde{\omega}'_k-\omega_j)\sim \Omega/N$ for this term, we obtain
the following estimate for the  width of the subradiant mode: 
$\tilde{\omega}''_k \approx \gamma$, where 
\begin{equation}
\label{width0}
\gamma \sim\tau\Omega^2/N^2.
\end{equation}
It can be seen that $\gamma$ is much smaller than the MFS (by the factor
 $\Omega\tau/N \ll 1$). This justifies neglecting 
$\omega''^2$ in the denominators of 
Eqs.\ (\ref{resys}) and (\ref{imsys}).
 
The superradiant  solution of Eqs.\ (\ref{resys}) and (\ref{imsys})
corresponds to the case $\omega''\gg \Omega$.
Then we readily obtain
$\tilde{\omega}'=N^{-1}\sum_j\omega_j$ and  
$\tilde{\omega}''=\Gamma\sim N/\tau$. 
We see that, indeed, $\Gamma/\Omega \sim N/\Omega\tau \gg 1$.
Therefore, 
{\em the superradiant band in the spectral intensity is not affected
by the disorder}.

We therefore conclude that cooperative emission is {\em not} destroyed
by  disorder. The spectrum of the system consists of a single 
superradiant  and $N-1$ subradiant eigenmodes. For large
$N/\Omega\tau$, the subradiant modes are well defined, since their
spectral widths are much smaller than the MFS.

\subsection{Limit of weak disorder}

In this subsection, we address a nontrivial question
about the fate of cooperative eigenmodes when the disorder
in frequencies vanishes.
In this limit, $\Omega\tau/N\rightarrow 0$,
all oscillator frequencies become equal, i.e.
$\omega_i\rightarrow\omega_0$. In the absence of cooperative coupling,
$\alpha_{ij}=0$ ($i \neq j$), the eigenfrequencies of the system are
those of individual oscillators with the energy width much larger than
the MFS, $1/\tau\gg \Omega/N$, so that the spectrum of the system is
degenerate.

However, the situation is more complicated in the presence of
cooperative coupling, $\alpha_{ij}\neq 0$. 
Consider the case of a ``point sample'', $\alpha_{ij}=1$. 
In this case, the width of subradiant modes is given by 
Eq.\ (\ref{width0}). Important is that although the MFS diminishes
with decreasing $\Omega$, the width $\gamma$ decreases
even faster: $\gamma/(\Omega/N)\sim \Omega\tau/N\rightarrow 0$. 
In orher words, in the presence of even a very weak disorder, 
the narrow subradiant peaks do {\em not} overlap. Therefore, the
cooperative modes remain {\em distinct} even 
though the ``bare'' oscillator modes were already degenerate. 

In the case of general coupling, the width $\gamma$ of
subradiant modes for small values of $\Omega\tau/N$ will be determined by
the {\em fluctuations} of $\alpha_{ij}$, as we will see below.

\subsection{Fluctuations of $\alpha_{ij}$}

Let us turn to the case with realistic coupling $\alpha_{ij}$. The
eigenfrequencies $\tilde{\omega}_k$ should now be determined from  
Eq.\ (\ref{eigen}), which in component form reads 
\begin{eqnarray}
\label{eigensys1}
\frac{1}{\tau}\sum_j
\frac{(\omega_j-\omega')(1+\sigma'_j)-\omega''\sigma''_j}
{(\omega_j-\omega')^2+\omega''^2}=0,\label{resys1}
\\
\frac{1}{\tau}\sum_j
\frac{\omega''(1+\sigma'_j)+(\omega_j-\omega')\sigma''_j}
{(\omega_j-\omega')^2+\omega''^2}=1,\label{imsys1}
\end{eqnarray}
with $\sigma_i(\omega)=\sigma'_i(\omega)+i\sigma''_i(\omega)$
satisfying Eq.\ (\ref{eqsigma}), or in component form 
\begin{eqnarray}
\label{sigsys1}
\sigma''_i+\frac{1}{\tau}\sum_j\, (\alpha_{ij}-1)\,
\frac{(\omega_j-\omega')(1+\sigma'_j)-\omega''\sigma''_j}
{(\omega_j-\omega')^2+\omega''^2}=0,\label{imsigsys1}
\\
\sigma'_i-\frac{1}{\tau}\sum_j \,(\alpha_{ij}-1)\,
\frac{\omega''(1+\sigma'_j)+(\omega_j-\omega')\sigma''_j}
{(\omega_j-\omega')^2+\omega''^2}=0.\label{resigsys1}
\end{eqnarray}
For  $\omega''\ll \Omega/N$, the system 
(\ref{eigensys1})--(\ref{resigsys1}) can be approximately
solved in the same way as for a ``point'' sample. 
The corresponding condition will be
derived in Section\  V. 
 
When evaluating the contribution to the lhs of Eq.\ (\ref{imsys1})
coming from the first term in the numerator, one should keep only
one term in the sum with $\omega_j$ closest to $\tilde{\omega}'_k$:
$(\omega_j-\tilde{\omega}'_k)\sim \Omega/N$. Then we obtain 
\begin{equation}
\label{width}
\tilde{\omega}''_k\sim \frac{\tau\Omega^2}{N^2}
\left(1-\frac{1}{\tau}\sum_j
\frac{\sigma''_j}{\omega_j-\tilde{\omega}'_k}\right),
\end{equation}
where we again dropped $\omega''^2$ in the denominator. 
Since $\sigma_i'\ll \sigma_i''\ll 1$ (see Section\  V), the frequencies
$\tilde{\omega}'_k$ in Eq.\ (\ref{width}) 
are the same as for the case 
 $\alpha_{ij}=1$. Finding $\sigma''_i$ in the
first order from Eq.\ (\ref{imsigsys1}), and substituting the result
into Eq.\ (\ref{width}), we obtain
\begin{equation}
\label{width1}
\tilde{\omega}''_k\sim \frac{\tau\Omega^2}{N^2}
\left[
1+\frac{1}{\tau^2}\sum_{ij}
\frac{\alpha_{ij}-1}
{(\omega_i-\tilde{\omega}'_k)(\omega_j-\tilde{\omega}'_k)}
\right].
\end{equation}
The second term is the sought correction to the width of the
subradiant modes. Remarkably, this term turns to {\em zero} if the
matrix elements $\alpha_{ij}$ are replaced by their average
$\bar{\alpha}$. Indeed, in this case the double sum in Eq.\ (\ref{width1})
would factorize into a product of two  sums, each vanishing
due to the fact that $\tilde{\omega}'_k$ are the solutions of 
Eq.\ (\ref{resys}) (corresponding to $\alpha_{ij}=1$). Therefore, the
widths of the subradiant modes are determined by the {\em fluctuations},
$\delta\alpha_{ij}$, of the coupling parameters $\alpha_{ij}$ rather
than the deviation of their average, $\bar{\alpha}$, from unity. 
It should be noted that  this property is general:
one can easily see by comparing Eqs.\ (\ref{resys1})\ and\  (\ref{imsys1}) 
to Eqs.\ (\ref{imsigsys1})\ and\  (\ref{resigsys1}) that for 
$\alpha_{ij}=const$, we have $\sigma''_i=0$ and
$\sigma'_i\ll 1$, so that the eigenfrequencies
$\tilde{\omega}_k$ are unaffected.

\subsection{Discussion of the numerical results}

We are now in the position to explain the spectra 
shown in Figs.\ 3--6. For the model coupling: $\alpha_{ij}=\alpha$,
$\alpha_{ii}=1$,  fluctuations only in the diagonal
 elements are finite: 
$\delta\alpha_{ij}=(1-\alpha)\delta_{ij}$. Substituting this
$\delta\alpha_{ij}$ into Eq.\ (\ref{width1}) [instead of
$(\alpha_{ij}-1)$] and keeping only the term
with $(\tilde{\omega}'_k-\omega_j)\sim\Omega/N$ in the remaining sum,
we obtain 
\begin{equation}
\label{modelwidth}
\gamma \sim \left[\frac{\tau\Omega^2}{N^2}+\frac{1}{\tau}(1-\alpha)\right].
\end{equation}
The above expression indicates that after the cooperative modes
have been formed (at $N\sim \Omega\tau$), the system can be found in 
two different  regimes characterized by the relative magnitude of
the first and second terms in the rhs. For intermediate number of
oscillators, 
$\Omega\tau\lesssim N \lesssim \Omega\tau\, (1-\alpha)^{-1/2}$, the
width decreaeses with increasing $N$, as can be seen by comparing
the bottom second and third curves in each of Figs.\ 3--6
(note that the lowest curves with $N=2$ show no sign of
cooperative emission). In this regime, the system
behaves in the same way as a ``point'' sample.
With increasing number of oscillators, the dependence 
on $N$ saturates, and  the width is dominated by the fluctuations of
$\alpha_{ij}$. Correspondingly, the change in the pattern of the peaks
in Figs.\ 3--6, calculated for different values of $(1-\alpha)$,
occurs at different $N$, as can be seen by comparing the next two
curves in each figure. Note, however, that with further increase 
in  $N$, the curves exhibit yet another change in
pattern. Namely, the peaks get smeared out (the top
two curves in each figure). This occurs when the value of
$(1-\alpha)/\tau$ exceeds the MFS, $\Omega/N$, which is
inconsistent with the  above analysis. The reason for such a
discrepancy is that for large $N$, the model with coupling 
$\alpha_{ij}$ independent of separation between oscillators,
becomes inadequate, as we mentioned above. 
For the correct description of the peaks smearing  at large $N$, the 
spatial dependence of $\alpha_{ij}$ is crucial;
this question is addressed in Section\  \ref{scaling}.
Nevertheless, for $N\lesssim \Omega\tau/(1-\alpha)$, this
model describes accurately the mesoscopic features of the spectral
intensity, as shown in the next section. 

\section{Strong Mesoscopics Regime}
\label{mesoscopic}

Let us now estimate the typical width of the radiation
eigenmodes due to the fluctuations in $\alpha_{ij}$. Since the
configurational average of the second 
term in Eq.\ (\ref{width1}) [with $\delta\alpha_{ij}$ instead of
$(\alpha_{ij}-1)$]  vanishes, we need to evaluate
$\langle(\tilde{\omega}''_k)^2\rangle$. Using the fact that only
diagonal terms in the average 
$\langle \delta\alpha_{ij}\delta\alpha_{i'j'}\rangle$  survive and
omitting the first term in Eq.\ (\ref{width1}), we write
\begin{equation}
\label{fluct}
\langle(\tilde{\omega}''_k)^2\rangle\sim
\left(\frac{\Omega^2}{\tau N^2}\right)^2
\left\langle
\sum_{ij}\frac{(\delta\alpha_{ij})^2}
{(\omega_i-\tilde{\omega}'_k)^2(\omega_j-\tilde{\omega}'_k)^2}
\right\rangle.
\end{equation}
The sum is dominated by the terms with
$(\omega_i-\tilde{\omega}'_k)\sim(\omega_j-\tilde{\omega}'_k)\sim \Omega/N$.
Since the typical spatial separation between two oscillators with
close frequencies is $\sim L$, the separation
fluctuations are of the same order. Thus, the typical fluctuation
of $\alpha_{ij}$ is
$\delta\alpha\equiv
\sqrt{\langle(\delta\alpha_{ij})^2\rangle}\sim \left(L/\lambda_0\right)^2$, and
we finally obtain the typical width of a subradiant mode,
$\gamma\equiv
\sqrt{\langle(\tilde{\omega}''_k)^2\rangle}$, as
\begin{equation}
\label{lastwidth}
\gamma\sim \frac{\delta\alpha}{\tau}
\sim\frac{1}{\tau}\!\left(\frac{L}{\lambda_0}\right)^2.
\end{equation}
Comparing Eq.\ (\ref{lastwidth}) to Eq.\ (\ref{width1}), we
see that fluctuations in $\alpha_{ij}$ dominate the width $\gamma$ 
for $N\gtrsim \Omega\tau(\lambda_0/L)$.

In order to characterize the fine structure in the emission
spectrum, it is convenient to introduce the dimensionless parameter 
\begin{equation}
\label{kappa}
\kappa=\frac{\Omega\tau}{N}\!
\left(\frac{\lambda_0}{L}\right)^2.
\end{equation}
It represents the product of a small,
$\Omega\tau/N$, and a large, $(\lambda_0/L)^2$, factors,
which characterize the disorder and the system size, respectively.
In terms of $\kappa$, the condition for the formation of the
cooperative modes, $N/\Omega\tau\gg 1$, can be presented as 
$\kappa\ll \left(\lambda_0/L\right)^2$.

Using Eq.\ (\ref{kappa}), the width (\ref{lastwidth}) can
be expressed  in terms of the MFS as
\begin{equation}
\label{rewritten}
\gamma\sim \frac{1}{\kappa}
\!\left(\frac{\Omega}{N}\right).
\end{equation}
This result applies when 
$\kappa\lesssim \lambda_0/L$. 
On the other hand, it was implicit in the above derivation 
(Section\  \ref{structure}) that typical 
$\sigma'_i$ and  $\sigma''_i$ are smaller than unity . The latter
parameters can be estimated in a similar way from 
Eqs.\ (\ref{imsigsys1})\ and\  (\ref{resigsys1}) with the result:
$\sigma''\sim \kappa^{-1}$ and 
$\sigma'\sim \sigma''^2 \sim \kappa^{-2}$.
Thus, the lower boundary for $\kappa$, at which Eq.\ (\ref{rewritten})
applies, is $\kappa\gtrsim 1$. For $\kappa \sim 1$, all terms in 
Eqs.\ (\ref{resys1})--(\ref{resigsys1}) become of the same order of
magnitude, and for smaller $\kappa$ this system has no 
{\em subradiant} solutions, as discussed above. 

Since the MFS exceeds the width $\gamma$ within the entire domain 
$1\lesssim \kappa \lesssim \lambda_0/L$, the fine structure
in the spectral intensity $I(\omega)$ is well pronounced.
In other words, this domain corresponds to the
strong mesoscopics regime. The opposite case $\kappa\lesssim 1$
is considered in the next section.

\section{Weak Mesoscopics Regime}
\label{scaling}

In the domain $\kappa \ll 1$, the system cannot sustain
eigenmodes that involve all $N$ oscillators. As a result, the
eigenmodes  become localized, in the sense that each eigenmode
would comprise some $N_c\ll N$ oscillators and occupy the volume
with characteristic size $L_c\ll L$. The magnitude of $L_c$ and $N_c$
can be estimated from the following argument. Let us divide
the system of oscillators into subsystems of increasingly 
smaller size. When the size of the subsystem becomes $\sim L_c$,
the system of equations (\ref{resys1})\ and\  (\ref{imsys1}),
{\em applied} {\em to} {\em a} {\em subsystem},
 first acquires a solution.
This happens when the width $\gamma_c\sim \tau^{-1}(L_c/\lambda_0)^2$,
determined from Eq.\ (\ref{lastwidth}) {\em for} {\em a} {\em subsystem},
becomes of the order of MFS  {\em within} {\em a} {\em subsystem},
i.e.
\begin{equation}
\label{main}
\frac{1}{\tau}\left(\frac{L_c}{\lambda_0}\right)^2
\sim \frac{\Omega}{N_c}.
\end{equation}
Taking into account that $N_c=N(L_c/L)^3$, we find
\begin{equation}
\label{c1}
L_c\sim\kappa^{1/5}L,\,\,\,\,\,\,\,N_c\sim\kappa^{3/5}N.
\end{equation}
Substituting these results back into Eq.\ (\ref{main}), we find for
the eigenmodes width 
\begin{equation}
\label{c2}
\gamma=\gamma_c\sim\frac{1}{\kappa^{3/5}}\left(\frac{\Omega}{N}\right)
\gg \frac{\Omega}{N}.
\end{equation}
From Eq.\ (\ref{c1}), we can also estimate how the relative  amplitude
of mesoscopic fluctuations in the spectral intensity  $I(\omega)$
falls off with decreasing $\kappa$:
\begin{equation}
\label{decreasing}
\frac{\delta I}{I}\sim \left(\frac{N_c}{N}\right)^{1/2}=\kappa^{3/10}.
\end{equation}
It is apparent that smearing  of the fine structure in the 
cooperative emission
spectrum with decreasing $\kappa$ occurs rather slowly.

\section{Dipole-dipole interactions}
\label{discussion}

In this section we study the effect of dipole-dipole
interactions on the cooperative emission from a disordered system.
Note that for a ``point'' sample with $L\ll \lambda_0$,
the  typical magnitude of the dipole--dipole interaction between
two oscillators is much larger than their superradiant coupling,
$\beta_{ij}/\alpha_{ij}\sim (\lambda_0/L)^3\gg 1$.
The structure of the eigenmodes in the absence of 
superradiant coupling, given by Eq.\ (\ref{lastsys}) 
with $\alpha_{ij}=0$, was considered in several
papers.\cite{lev,akulin,bur,sto1,sto2,par}
Renormalization--group arguments of  
Ref. \onlinecite{lev} (see also Ref. \onlinecite{bur})
suggest that all eigenmodes are delocalized.
Numerical studies\cite{sto1,sto2,par} 
indicate a wide range of spatial scales in
eigenmodes and, thus, seem to support this conclusion.
In Ref. \onlinecite{akulin}, the role of general random-matrix
perturbation in the spectrum of multilevel system was studied
analytically; he ensemble-averaged renormalization of the spectrum of
the system was derived which does not capture, however, the
mesoscopic effects. Below we argue that finite disorder in combination
with superradiant coupling lead to a certain ``resistance''
of the system to  large, but zero on average, dipole--dipole terms
because of the formation of cooperative modes.

In the absence of superradiant coupling ($\alpha_{ij}=0$)  the
dipole--dipole interactions lead  to the shifts 
in  the frequencies of individual oscillators. The resulting additional
spread in $\omega_i$ is, in general, much larger than
the ``bare'' spread $\Omega$. This can be readily seen from the
lowest--order correction to the frequency, $\delta\omega_i$, which
has the form 

\begin{equation}
\label{oscshift}
\delta\omega_i=
\frac{1}{\tau^2}\sum_{j\neq i}\frac{\beta_{ij}^2}{\omega_i-\omega_j},
\end{equation}
(since $\beta_{ii}=0$, the lowest--order correction to $\omega_i$ is
quadratic). The main contribution to the sum comes from 
pairs of oscillators located closely in space, with $r_{ij}\sim LN^{-1/3}$
(nearest neighbor interaction), so that 

\begin{equation}
\label{nearest}
\beta_{ij}\sim N\left(\frac{\lambda_0}{L}\right)^3.
\end{equation}
Since the typical frequency difference
for such pairs is $\sim \Omega$, we obtain 
\begin{equation}
\label{oscshift1}
\delta\omega_i\sim \frac{N^2}{\Omega\tau^2}
\left(\frac{\lambda_0}{L}\right)^6.
\end{equation}
On the other hand, the results obtained in the previous sections
apply only if the additional disorder, caused by  dipole--dipole
interactions, does not affect the MFS. This requires the condition 
$\delta\omega_i\ll\Omega$ to be met. Using Eq.\ (\ref{oscshift1}),
this condition could be rewritten as
$\lambda_0/L\ll (\Omega\tau/N)^{1/3}$. Since the formation
of cooperative modes occurs only if $\Omega\tau/N \ll 1$, one could
draw the conclusion that neglecting the dipole--dipole interactions
would be inconsistent with our basic assumption $L\ll \lambda_0$.

The resolution of this apparent contradiction lies in the
observation that,  in the presence of superradiant coupling,
i.e. $\alpha_{ij}\neq 0$, the true eigenmodes of the 
system are cooperative modes comprised from a large number of
oscillators. Therefore, the relevant condition should involve 
the shifts, $\delta\tilde{\omega}'_k$, of the {\em eigenmodes
frequencies}, rather than $\delta\omega_i$.
In the first order, $\delta\tilde{\omega}'_k$ is given by an
expression similar to the second
term in the rhs of Eq.\ (\ref{width1}) [with $\beta_{ij}$ instead of
$(\alpha_{ij}-1)$]. Since  this term {\em vanishes} on
average, as discussed above, the typical shift, 
$\delta\tilde{\omega}'
\equiv\sqrt{\langle(\delta\tilde{\omega}'_k)^2\rangle}$, 
can be estimated from  [compare with Eq.\ (\ref{fluct})] 
\begin{equation}
\label{dipole}
\langle(\delta\tilde{\omega}'_k)^2\rangle\sim
\left(\frac{\Omega^2}{\tau N^2}\right)^2
\left\langle
\sum_{ij}\frac{(\beta_{ij})^2}
{(\omega_i-\tilde{\omega}'_k)^2(\omega_j-\tilde{\omega}'_k)^2}
\right\rangle.
\end{equation}
There are two main contributions to the sum in the rhs.
The first comes from the nearest--neighbor
interaction with  $\beta_{ij}$ given by Eq.\ (\ref{nearest}). The
second contribution originates from the pairs ($ij$) which are close 
in frequency; for such pairs, $\beta_{ij}\sim (\lambda_0/L)^3$. 
Both contributions turn out to be of the 
{\em same} order of magnitude, resulting in
\begin{equation}
\label{eigenshift}
\delta\tilde{\omega}'\sim \frac{1}{\tau}
\left(\frac{\lambda_0}{L}\right)^3.
\end{equation}
This result is smaller than $\delta\omega_i$ in 
Eq.\ (\ref{oscshift1})  by the factor
$N(N/\Omega\tau)(\lambda_0/L)^3\gg 1$. Such a dramatic difference
illustrates the ``resistance'' of a coupled system
of oscillators with disorder in frequencies to  dipole--dipole
interactions, as mentioned above. This property can also be
qualitatively explained as follows. 
The dipole--dipole interaction between two subradiant modes
can be viewed as an interaction between a mode and the electric field,
$\tilde{\bf E}({\bf r})$,
created by the dipole moments of oscillators making up
the other mode. Since the number of oscillators in a mode is large,
their electric fields effectively cancel each other, so that the
resulting net field, $\tilde{\bf E}({\bf r})$, varies in space
{\em much slower} than those of the individual 
oscillators. Note now that a slowly varying electric
field couples only weakly to a {\em subradiant} mode. 
In fact, the  suppression of the dipole--dipole interaction between
subradiant modes has the same physical origin as their decoupling 
from the radiation field: had the electric
field $\tilde{\bf E}$ been uniform, the cooperative
modes would not interact at all with each other. 
This is the reason why the corrections to  
$\tilde{\omega}'_k$ and $\tilde{\omega}''_k$ 
vanish {\em on average}, and consequently the 
typical $\delta\tilde{\omega}'$
and $\gamma$ are determined by the {\em fluctuations} of
$S_{ij}$.  In contrast, the frequency shifts of individual
oscillators are due to their interactions with the nearest
neighbors, so that no cancellations occur.

Thus we arrive at the condition
$\lambda_0/L \ll \left(\Omega\tau\right)^{1/3}$, 
or, in terms of the parameter $\kappa$, 
\begin{equation}
\label{condition}
\kappa\gg \frac{1}{N}\left(\frac{\lambda_0}{L}\right)^5.
\end{equation}
This condition should be consistent with the condition
for the formation of the cooperative modes, 
$\kappa\ll (\lambda_0/L)^2$. We see that both conditions are
satisfied for sufficiently large  $N$, i.e. $N\gg (\lambda_0/L)^{3}$.
To account for different mesoscopics regimes, 
it is convenient to present  Eq. (\ref{condition}) in the form  
\begin{equation}
\label{regime}
N\gg \left(\frac{\lambda_0}{L}\right)^n.
\end{equation}
Then $n=3$,\ 4,\ and\ 5 correspond to the ``point'' sample,
strong mesoscopics, and weak mesoscopics regimes, respectively.

\section{Conclusions}
\label{conclusion}

The main result of the present paper is that  disorder
in oscillators frequencies   does not destroy the cooperative
character of the emission from a ``point'' sample, as long as the
MFS, $\Omega/N$, is smaller than the linewidth of
an individual oscillator, $\tau^{-1}$. In the opposite case, when 
$\Omega/N \gg \tau^{-1}$, the
spectrum represents a system of non--overlapping Lorentzians with
the width $\tau^{-1}$.

It is convenient to characterize the disorder 
in terms of the dimensionless parameter 
$\kappa=\Omega\tau\lambda_0^2/NL^2$.
Below we summarize our results for the
characteristic width, $\gamma$, of the subradiant  peaks
(in  units of $\Omega/N$) for different domains of  $\kappa$:
\begin{equation}
\label{domains}
\gamma=\frac{\Omega}{N}\,\Phi\left(\kappa,L/\lambda_0\right),
\end{equation}
where the dimensionless function $\Phi$ has the following
asymptotes
\begin{eqnarray}
\label{asymptotes}
\Phi&=&\kappa\!\left(\frac{L}{\lambda_0}\right)^2,
\,\,\,
{\rm for}
\,\,\,
\frac{\lambda_0}{L}
\lesssim\kappa\lesssim\left(\frac{\lambda_0}{L}\right)^2,
\,\,
\mbox{``point''\ sample},
\nonumber\\
\Phi&=&\kappa^{-1},
\,\,\,\,\,\,\,\,\,\,\,\,\,\,\,\,
{\rm for}
\,\,\,
1\lesssim\kappa\lesssim \frac{\lambda_0}{L},
\,\,
\mbox{strong\ mesoscopics},
\nonumber\\
\Phi&=&\kappa^{-3/5},
\,\,\,\,\,\,\,\,\,\,\,\,
{\rm for}
\,\,\,
\kappa\lesssim 1,
\,\,
\mbox{weak\ mesoscopics}.
\end{eqnarray}
For $\kappa\gtrsim (\lambda_0/L)^2$, the spectrum corresponds to
uncoupled oscillators. 

In Section\ \ref{simple}, we presented an exact solution of a
model with simplified (separable) coupling Eq.\ (\ref{separab}).    
This model describes accurately the first two (``point'' sample and
strong mesoscopics) regimes in Eq.\ (\ref{asymptotes}). It
becomes, however, inadequate in the third (weak mesoscopics) regime, giving
a $\kappa^{-1}$ instead of the correct $\kappa^{-3/5}$ dependence
for the period of mesoscopic structure in the cooperative emission
spectrum. 

Throughout the paper we have considered a three-dimensional  system
of oscillators. When the oscillators are confined to a plane,
only the results for $\kappa \lesssim 1$ should be modified.
In this case, repeating the consideration of Section\ \ref{scaling}, we
obtain $\Phi=\kappa^{-1/2}$. Also for the relative magnitude of
mesoscopic fluctuations, $\bigl(\delta I/I\bigr)$, instead of 
Eq. (\ref{decreasing})
we obtain $\bigl(\delta I/I\bigr)\sim \kappa^{-1/4}$.

Note finally that in experiments, such as photoexcited excitons
in polymer films, the number of  oscillators $N$ is governed by the
excitation intensity.\cite{varde} Thus, for a given disorder, the
crossover from the strong mesoscopics regime ($\kappa > 1$) 
to the weak mesoscopics regime ($\kappa < 1$)
can be simply achieved by increasing the excitation intensity level.\cite{muk}

\acknowledgments

The authors are grateful to S. Mukamel and M. I. Stockman
for helpful discussions. The work at Vanderbilt University was
supported by  NSF grant ECS-9703453, and the work at the 
University of Utah was supported by NSF grant
DMR-9732820. M.E.R. was also supported by the Petroleum Research
Fund under grant ACS-PRF\# 34302-AC6.

\begin{figure}
\caption{Spectral intensity of $N$ identical oscillators calculated
from Eq.\ (\ref{N})  plotted vs.
$\Delta\omega=\omega-\omega_0$ for
$N=10$, and 
$\alpha=0$ (long--dashed line), $\alpha=0.5$ (dashed line), 
$\alpha=0.8$ (dotted line), and $\alpha=0.9$ (solid line).}
\end{figure}
\begin{figure}
\caption{{\em Uncoupled oscillators}: Spectral intensity $I(\omega)$
plotted vs. $\Delta\omega=\omega-\omega_0$ for
several sets of random oscillator frequencies
with $\Omega\tau=5.0$ and  
$\alpha=0$.}
\end{figure}
\begin{figure}
\caption{{\em Coupled oscillators}: Spectral intensity
$I(\omega)$ calculated from Eq.\ (\ref{Iav}) for several sets of
random oscillator frequencies
with $\Omega\tau=5.0$ and $\alpha=0.8$.}
\end{figure}
\begin{figure}
\caption
{Same as in Fig. 3, but for $\alpha=0.85$.}
\end{figure}
\begin{figure}
\caption
{Same as in Fig. 3, but for $\alpha=0.9$.}
\end{figure}
\begin{figure}
\caption
{Same as in Fig. 3, but for $\alpha=0.95$.}
\end{figure}

\clearpage

\epsfxsize=6.0in
\epsffile{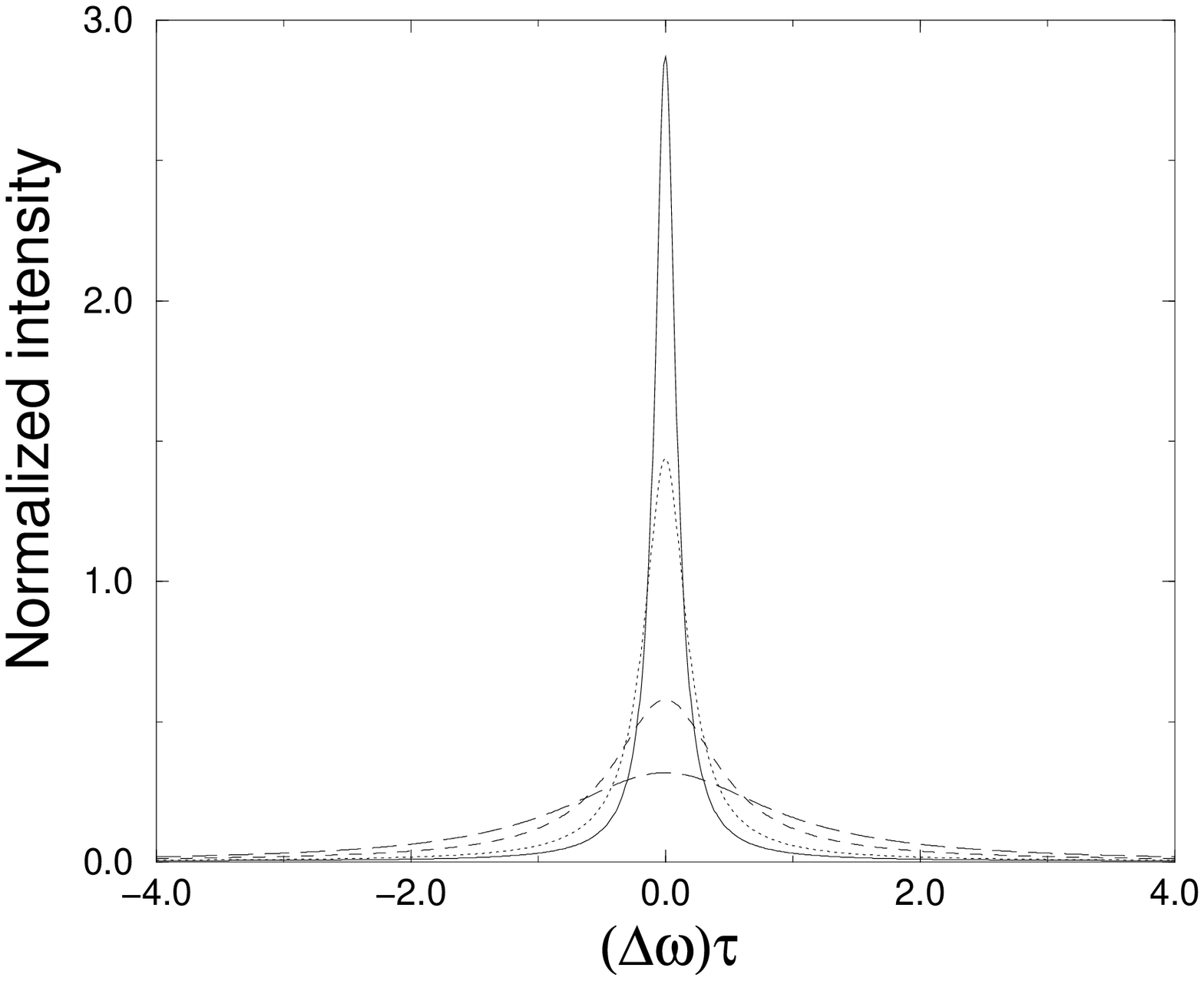}
\vspace{80mm}
\centerline{FIG. 1}
\epsfxsize=6.0in
\epsffile{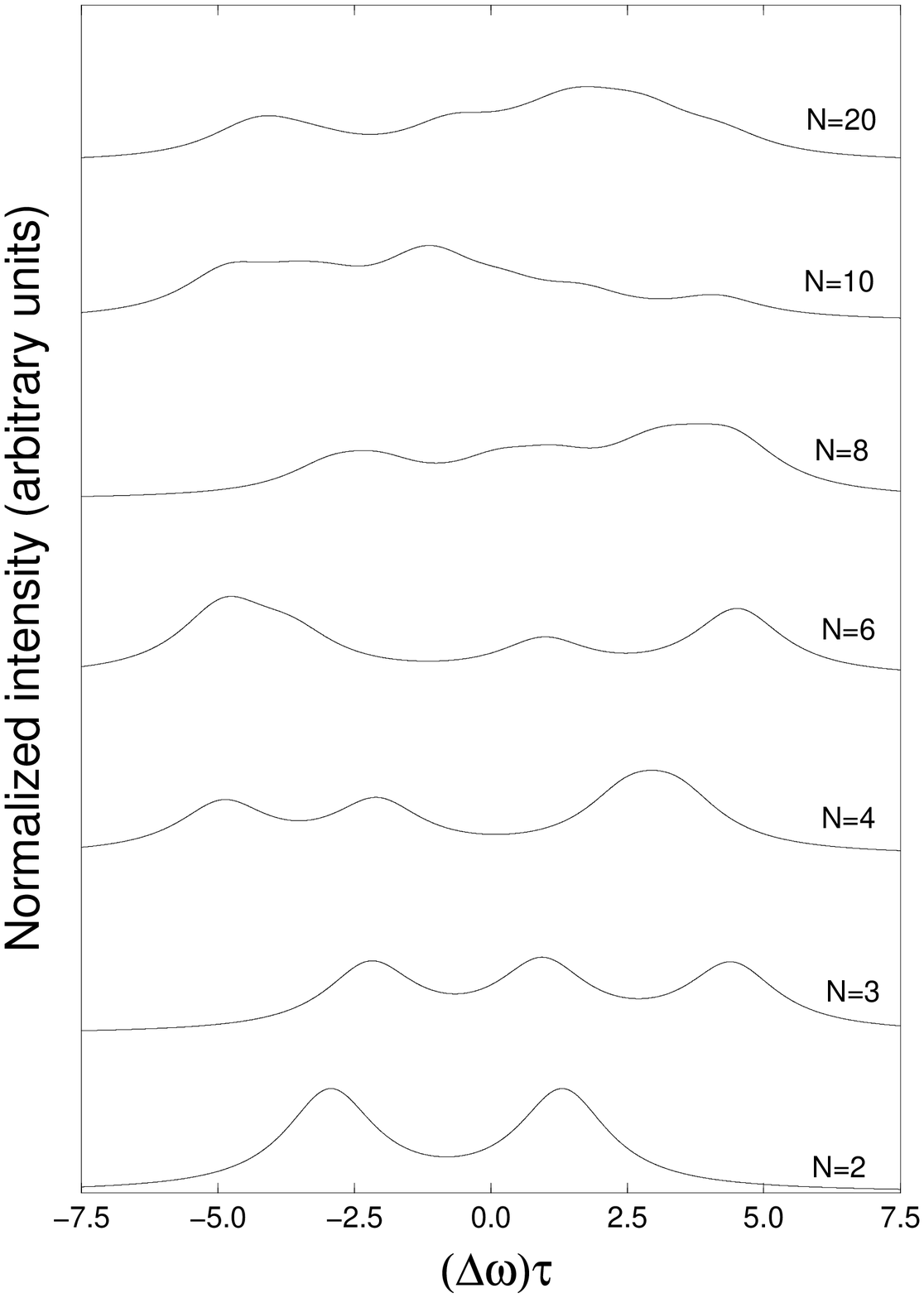}
\vspace{10mm}
\centerline{FIG. 2}
\epsfxsize=6.0in
\epsffile{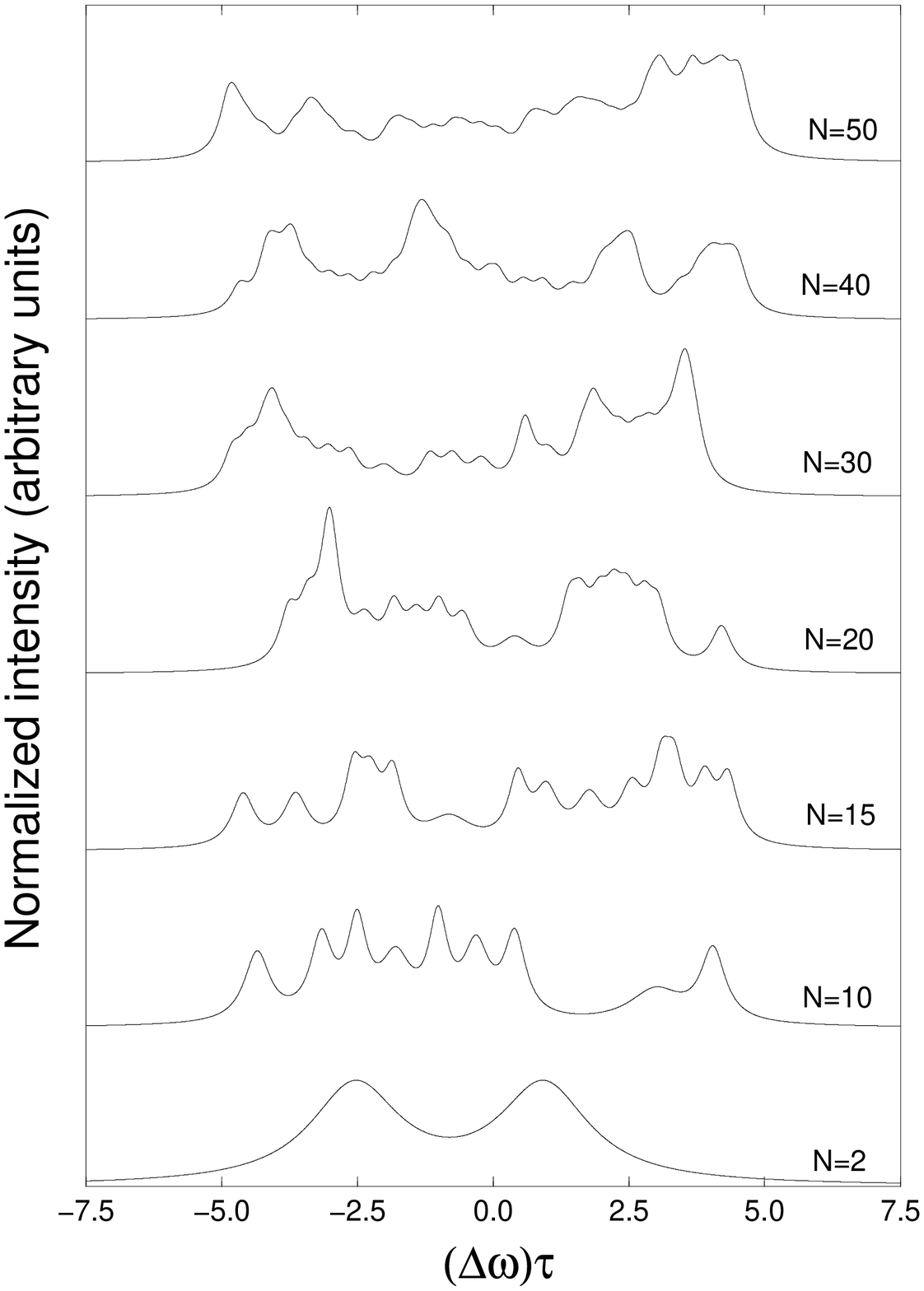}
\vspace{10mm}
\centerline{FIG. 3}
\epsfxsize=6.0in
\epsffile{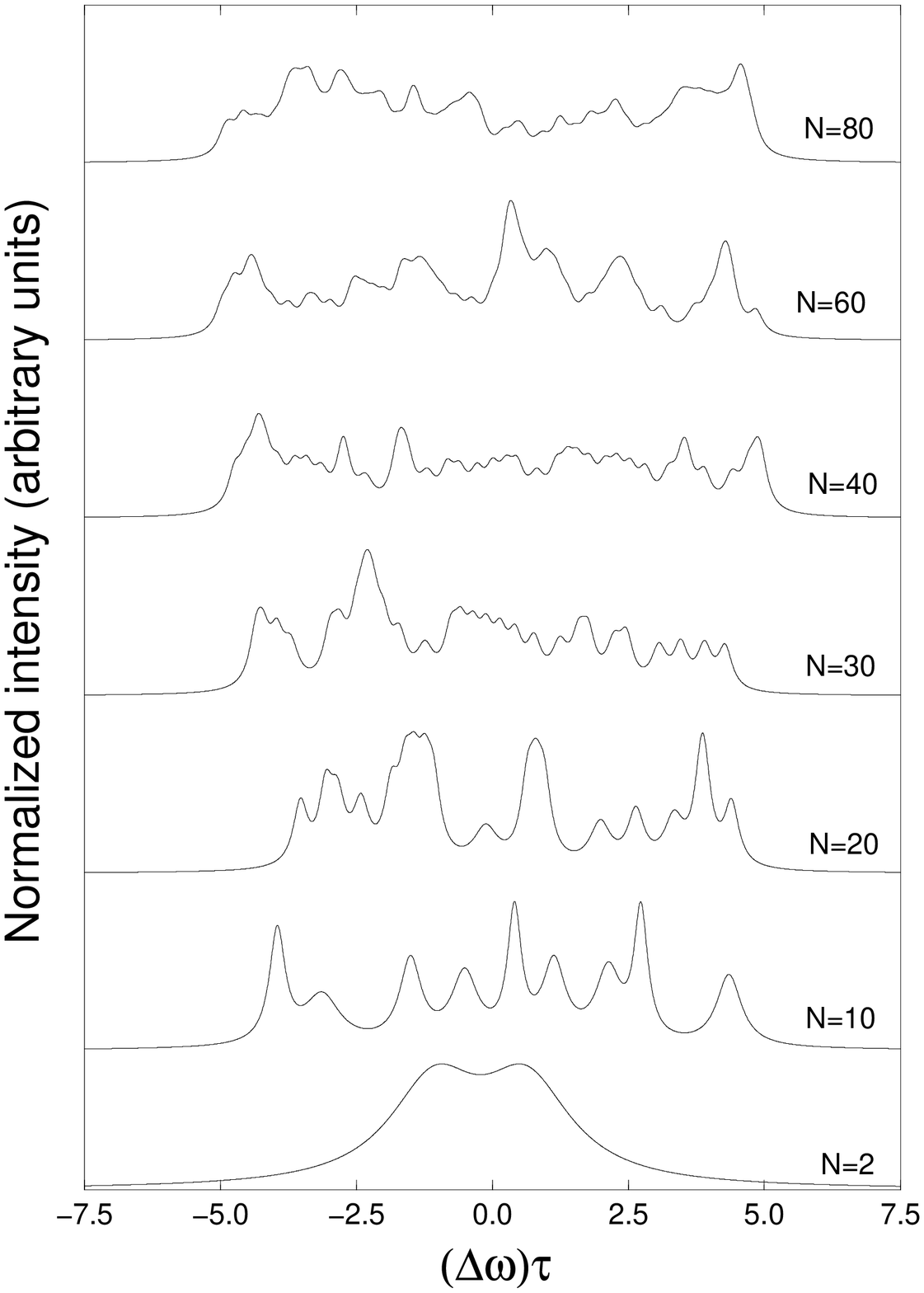}
\vspace{10mm}
\centerline{FIG. 4}
\epsfxsize=6.0in
\epsffile{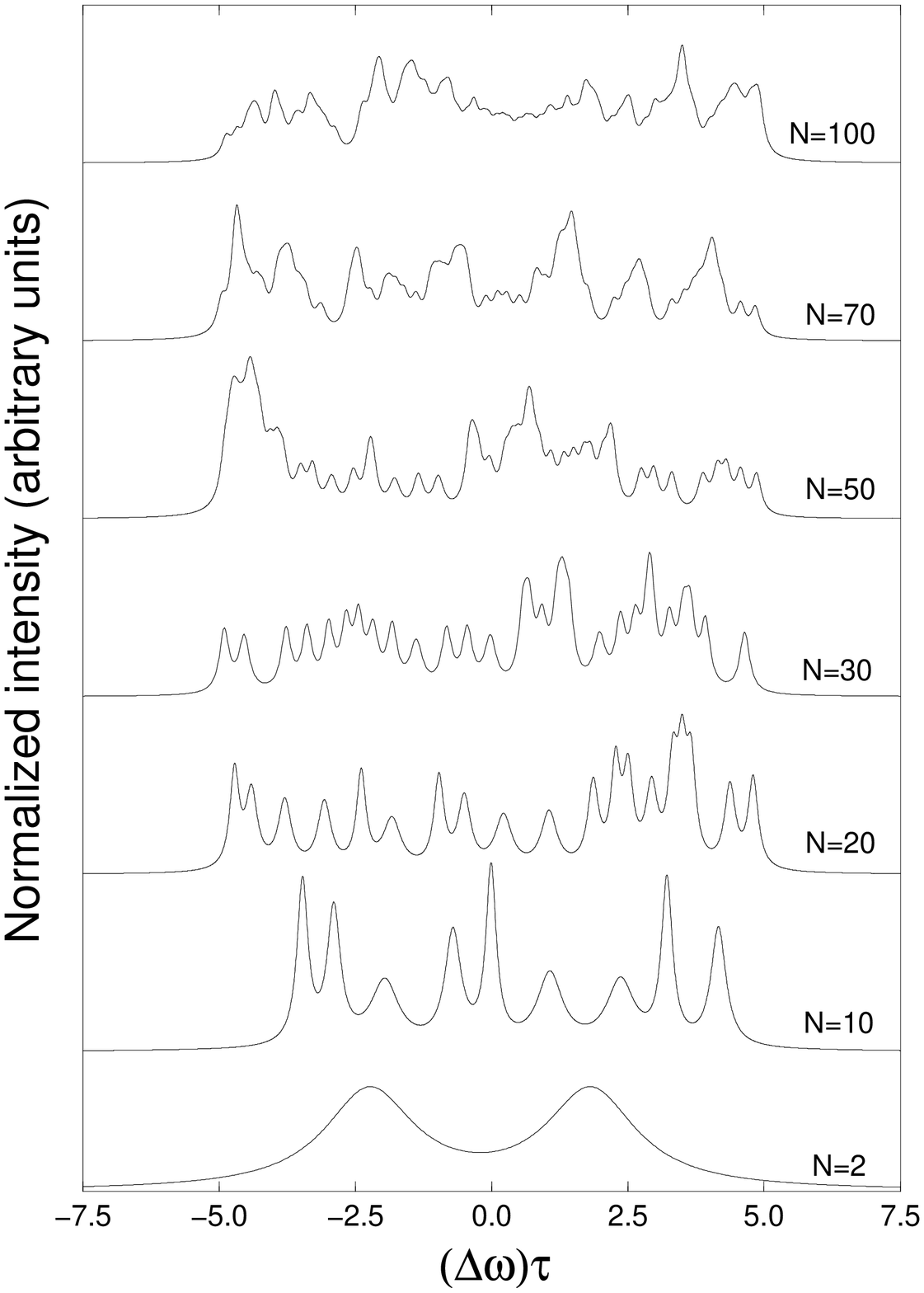}
\vspace{10mm}
\centerline{FIG. 5}
\epsfxsize=6.0in
\epsffile{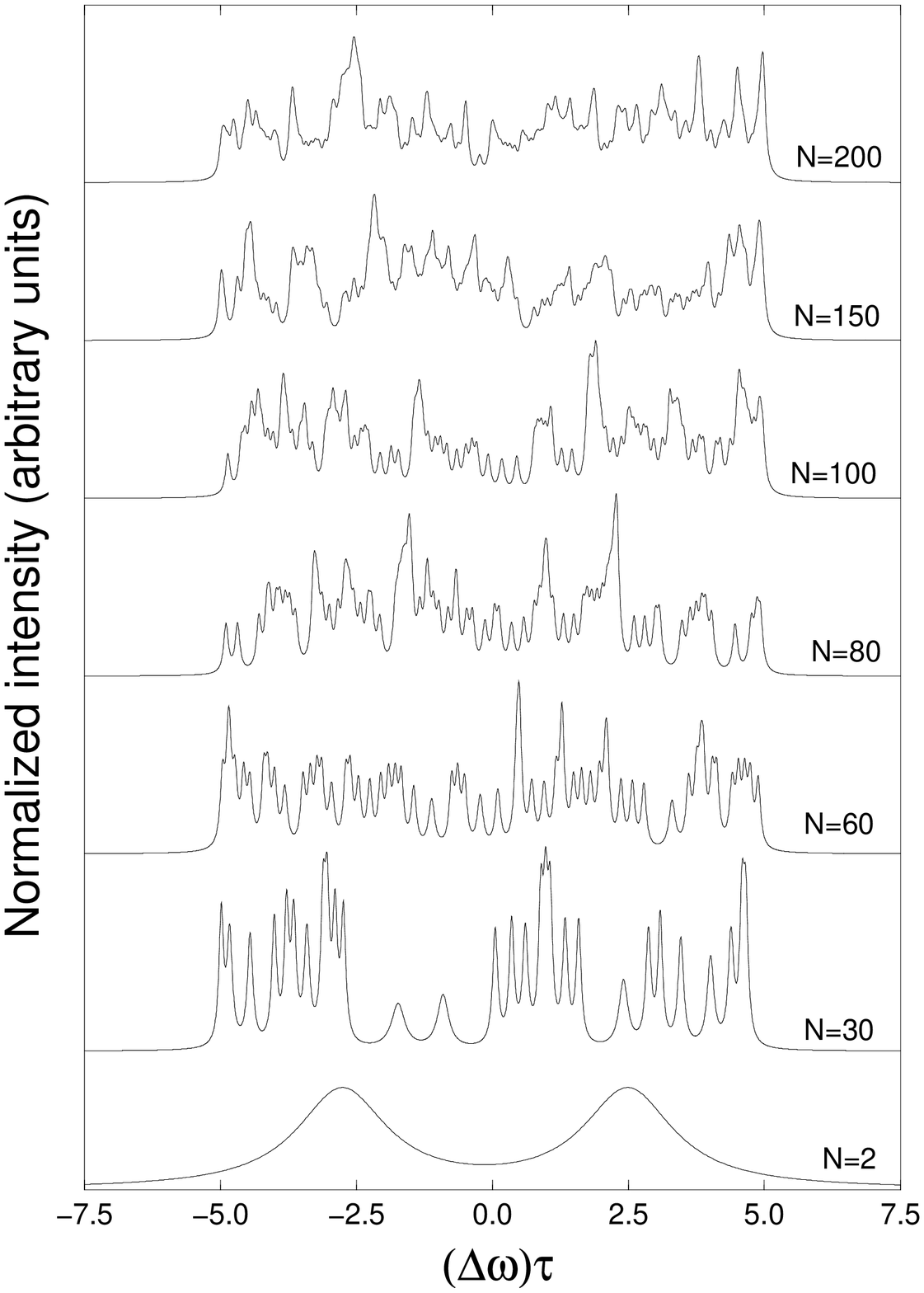}
\vspace{10mm}
\centerline{FIG. 6}

\end{document}